\documentclass[preprint]{emulateapj}

\newcommand\beq{\begin{equation}}
\newcommand\eeq{\end{equation}}
\newcommand\beqar{\begin{eqnarray}}
\newcommand\eeqar{\end{eqnarray}}

\newcommand\etal{{et al.~}}

\newcommand{\flf}{\mathcal{F}}
 
\begin{document}

\title{The Spectral Index Distribution of EGRET Blazars: Prospects for
GLAST}

\author{Tonia M. Venters\altaffilmark{1} \& 
Vasiliki Pavlidou\altaffilmark{2,3}}

\altaffiltext{1}{Department of Astronomy and Astrophysics, The University 
of Chicago, Chicago, IL 60637}
\altaffiltext{2}{
Kavli Institute of Cosmological Physics,
The University of Chicago, Chicago, IL 60637}
\altaffiltext{3}{Enrico Fermi Institute, The University of Chicago, Chicago, IL
60637}

\begin{abstract}
The intrinsic distribution of spectral indices in GeV energies 
of gamma-ray--loud blazars is
a critical input in determining the spectral shape of the unresolved
blazar contribution to the diffuse extragalactic gamma-ray background,
as well as an important test of blazar emission theories. 
We present a maximum-likelihood 
method of determining the intrinsic spectral index 
distribution (ISID)
of a population of $\gamma$-ray emitters which accounts for
error in measurement of individual spectral indices, and we apply it
to EGRET blazars. We find that the most likely Gaussian
ISID for EGRET blazars has a mean of $2.27$ and a standard deviation of
$0.20$. 
We additionally find some indication that FSRQs and BL Lacs may 
have different ISIDs (with BL Lacs being harder). 
We also test for spectral index hardening associated with blazar
variability for which we find no evidence. 
Finally, we produce simulated GLAST spectral index
datasets and perform the same analyses.  With improved statistics due to the much larger number of resolvable
blazars, GLAST data will help us determine the ISIDs with much
improved accuracy. Should any difference exist between the ISIDs of BL
Lacs and FSRQs or between the ISIDs of blazars in the quiescent and 
flaring states, GLAST data will be adequate to separate these ISIDs at a
significance better than $3\sigma$. 
\end{abstract}


\maketitle 

\section{Introduction}

The {\it Energetic Gamma-Ray Experiment Telescope} (EGRET) aboard the {\it Compton Gamma-Ray Observatory} 
observed the gamma-ray sky in
energies between ${\rm 30 \, MeV}$ and $\sim 10 {\rm \,  GeV}$ between 1991
and 2000. The third (and last) catalog of 
 EGRET point sources (Hartmann \etal 1999) included 271 resolved
 objects of which 93 were identified, either confidently or potentially, as blazars (gamma-ray loud active galactic
 nuclei). Thus, blazars constitute the class of $\gamma$-ray emitters
 with the largest number of identified members. 

The term ``blazar'' is used to refer collectively to BL Lac objects
and $\gamma-$ray loud flat spectrum radio quasars (FSRQs). 
Blazars are believed to be active galactic nuclei (AGNs) with the jet aligned
with our line-of-sight (Blandford \& K\"{o}nigl 1979; see 
Urry \& Padovani 1995 for a review of
general properties of blazars).
The main contribution to the $\gamma$-ray
emission of blazars is generally thought to result from either inverse Compton
scattering by relativistic electrons in the jet of lower energy
photons, produced either by synchrotron emission within the jet (SSC), or by emission external to the jet, for example from the accretion disk
around the central engine (EC);
or from proton-induced cascades (see, e.g., von Montigny et al.\ 1995, 
B\"ottcher 2006 and references therein
for a review of blazar $\gamma-$ray emission processes).

Blazars fainter than the ones in the 3rd EGRET (3EG) catalog (and thus 
unresolved by EGRET) are expected to have a sizable 
contribution to the diffuse, isotropic gamma-ray background detected by EGRET
(Sreekumar \etal 1998) which is presumably of extragalactic origin. 
The exact amount of this contribution remains unclear, 
as different models for the blazar $\gamma$-ray luminosity function 
result in very different predictions for the cumulative $\gamma$-ray flux
from all unresolved blazars, ranging from a few to $100\%$ of the
EGRB (Padovani \etal 1993; Stecker \etal 1993; Salamon \& Stecker 1994; Chiang
\etal 1995; Stecker \& Salamon 1996a, hereafter SS96a; Kazanas \& Perlman 1997; 
Chiang \& Mukherjee 1998; Mukherjee \&
Chiang 1999; M\"{ucke} \& Pohl 2000; Kneiske \& Mannheim 2005; Dermer 2007;
Giommi \etal 2006; Narumoto \& Totani 2006). 
Such uncertainties aside, blazars must be accounted for in any attempt to understand, model, or otherwise make use of the
high-energy diffuse photon inventory (e.g. in constraining exotic high-energy
physics). The lowest
reasonable value we can expect for the cumulative diffuse intensity of
unresolved faint blazars yields a strong lower limit for the level of the
extragalactic gamma-ray background (EGRB).  
To constrain contributions from any other plausible
source, this lower limit must be subtracted from the observed background. 

A second critical property of blazars is their spectral index
distribution (SID). The energy spectrum of the $\gamma$-ray emission of blazars
in the EGRET energy range can be well approximated by a power law,
$F(E) \propto E^{-\alpha}$ with values of $\alpha$, the spectral index, for
individual blazars fitted to be in most cases between $2$ and $3$. 
Blazar spectra in the $\gamma$-ray regime encode important
information concerning particle acceleration and emission processes 
in blazar jets (see von Montigny \etal 1995 and references therein). In addition,  the distribution of 
blazar $\gamma$-ray spectral indices is a critical input in the
estimation of the unresolved blazar contribution to the extragalactic 
diffuse background (SS96a; Pohl \etal 1997).  
Whether the collective emission of unresolved blazars may be the
dominant component of the diffuse extragalactic gamma-ray background
depends not only on 
intensity, but also on spectral shape (Stecker \& Salamon 1996a,b; Pohl \etal
1997; Strong \etal 2004).  Additionally, blazars may 
constitute the dominant component of the extragalactic gamma-ray background
only at certain energies (Pavlidou \& Fields
2002).  In order to assess these issues, the spectral features of the
unresolved blazar emission need to be studied, and hence the spectral index
distribution of blazars needs to be understood. 

\subsection{Past Work}

Obtaining the spectral index distribution of blazars presents
three major difficulties:

 (1) Large measurement uncertainties in individual blazar spectral indices, due to low photon statistics.
These errors contaminate the sampling of the underlying {\em
intrinsic} spectral index distribution (ISID)\footnote{In this paper, 
we will use the term
  ``spectral index distribution'' (SID) to refer to the distribution
  of measured spectral indices and the term ``intrinsic spectral index
distribution'' (ISID) to refer to the true distribution of spectral
indices of the blazar population (i.e. free of any contamination due
to measurement errors).}, by exaggerating its spread, 
possibly to a large degree.

(2) Possible systematic change of the spectral index with flaring 
[suggested, e.g., by von Montigny \etal 1995; 
Mukherjee \etal 1996 (however, note that for the case of PKS 0528+134,
 the significance of the result was reduced by subsequently obtained
data, see Mukherjee \etal 1997); SS96a;  
M\"{ucke} \etal 1996; Pohl \etal 1997]. 

(3) Possible existence of two spectrally distinct populations 
(BL Lacs and FSRQs) in the resolved blazar sample 
(e.g. Mukherjee \etal 1997; Pohl \etal 1997) 

Despite these difficulties, several authors have studied
different aspects of the statistical properties of GeV blazar spectra. 
SS96a calculated a spectral index distribution
for resolved EGRET blazars which they then used to derive the spectral 
shape of the collective emission from unresolved blazars. They
recognized that an appreciable spread in blazar spectral indices will
lead to a pronounced concavity in the collective emission spectrum
(see Brecher \& Burbidge 1972), a
feature which is also at least tentatively present in determinations 
of the EGRB based on EGRET data (Sreekumar
\etal 1998; Strong \etal 2004). Additionally, they insightfully 
emphasized the potential importance of both variability as well as 
measurement errors in the determination of the blazar SID.
SS96a remains to this day the only work
associating the distribution of blazar spectral indices as measured in
individual objects with a model predicting the spectral shape of the
unresolved blazar emission. 

However, their treatment suffers from three major unresolved
problems. First, their treatment of variability 
was based on very uncertain
information from very few objects. 
Second, BL Lacs and
FSRQs are treated as a single population, while the validity of this
assumption was not evaluated.  Finally,  
their treatment of measurement errors 
worsens, rather than alleviates, 
the overestimation of the SID
spread, and thus,
overestimates the curvature of the unresolved blazar emission spectrum.
This problem will {\em not} 
automatically disappear when the much larger sample of detected blazars 
and corresponding measured spectral indices of the upcoming GLAST
mission becomes available. The bulk number of blazar detections will
always be close to the instrument sensitivity limit, and will involve
only a few tens of photons from each object, which means that the bulk
number of measured spectral indices will always have substantial
measurement errors associated with them. Therefore, in order to be able to
utilize all  future measurements, including those for faint objects,
in understanding the spectral properties of $\gamma$-ray--loud AGN in the GeV
energy range, spectral index uncertainties have to be carefully dealt with. 

Pohl \etal (1997) used a different method to derive the spectral shape
of the collective unresolved emission of blazars, which effectively
circumvents the problem of large uncertainties in the measurement of
spectral indices of individual objects. They co-added the spectra of
resolved AGN and pointed out that if the spectral properties of unresolved
blazars are similar to those of resolved blazars, then 
this co-added spectrum is the most appropriate quantity (in terms
of spectral shape) for comparison with observations of the
EGRB. They performed the analysis separately for BL Lacs and
FSRQs, and found that the BL Lac co-added spectrum is harder by $\delta \alpha
= 0.12 \pm 0.08$ than that of FSRQs, although the significance of
the difference is low due to the low-number BL Lac statistics. Similarly, they
found a difference between the FSRQ co-added flaring spectrum and the
time-averaged co-added FSRQ spectrum, with the flaring spectrum being harder by
$\delta \alpha = 0.18 \pm 0.05$, a result with a statistical significance
between $3$ and $4\sigma$. They verified that the co-added spectra are
concave, although their method yields a much smaller spectrum curvature than
the SS96a model. 

Mukherjee \etal (1997), as part of a comprehensive analysis of the EGRET
blazar observations available at that time, evaluated the {\em average}
spectral properties of the blazar population and tested for indications of
evolution of the spectral index with redshift and spectral differences between
FSRQs and BL Lacs. Assuming equal errors in individual measurements of
spectral indices, they found an average spectral index of $2.15\pm0.04$ for
all blazars, and average spectral indices of $2.03\pm0.09$ and $2.20\pm0.05$
for the subsets of BL Lacs and FSRQs, respectively. They concluded that the
statistical significance of the difference between the mean spectral indices
of the two populations was at the level of $2.5\sigma$ and was not enough to
justify a spectral separation of the two populations.  Finally, they tested for
redshift evolution of blazar spectra and found that for both BL Lacs and
FSRQs, the data were consistent with no evolution. 
Neither Pohl \etal (1997) nor Muhkerjee et al. (1997) derived an SID
for 
blazars. 

\subsection{This Work}

The {\it Gamma-ray Large Area Space
  Telescope} (GLAST), which is scheduled for launch in 2007, will be able to address 
and ameliorate these difficulties. It will detect many
more blazars than EGRET (between $1,000$ and $10,000$ blazars;
Stecker \& Salamon 1999; Narumoto \& Totani 2006;
Dermer 2007), enabling confirmation or rejection of any statistical trends seen in
EGRET data. In order to facilitate the spectral studies of blazars with GLAST,
it is important to use EGRET data to identify open questions and
issues which can benefit from the dramatically improved GLAST
statistics. Furthermore, it is advantageous to use EGRET data to identify and
test appropriate statistical techniques which will allow us to
deal with observational difficulties which are also expected to be present in
GLAST data (such as large, varying errors of measurement in individual
spectral indices).

In this context, with this paper we attempt to:
(a) assess whether there is statistically significant evidence 
for evolutionary effects on the spectral index, manifesting as a 
correlation between the spectral index with either redshift or (isotropic)
photon luminosity (\S \ref{EVOL});
(b) estimate the extent to which individual measurement 
errors affect the sampling of the SID of EGRET blazars (\S \ref{efer}) and
perform a maximum likelihood analysis  
which accounts for these errors and determines the ``most likely'' parameters of the ISID (\S \ref{LIKEL}); 
(c) assess whether there is statistically significant evidence for a
    difference between the ISIDs of BL Lacs and FSRQs (\S \ref{BvsF})
    and the ISIDs of flaring and quiescent blazars (\S
    \ref{VARB});
(d) predict to what extent the upcoming GLAST observations will 
help us in resolving the issues mentioned above (\S \ref{GLAST}).
We summarize and discuss our conclusions in \S \ref{Disc}. 

The dataset used for our analysis, except where explicitly stated otherwise,
is the set of 66 blazars characterized as ``confident AGN
identifications'' in the 3EG catalog. We divide the population into 14
BL Lacs and 51 FSRQs (four sources have not been used due to lack of
measured P1234 fluxes) as in Nolan \etal (2003). The spectral indices of individual objects are the average, P1234 indices quoted in 3EG.

\section{Does the blazar SID depend on redshift or luminosity?}
\label{EVOL}

\begin{figure}
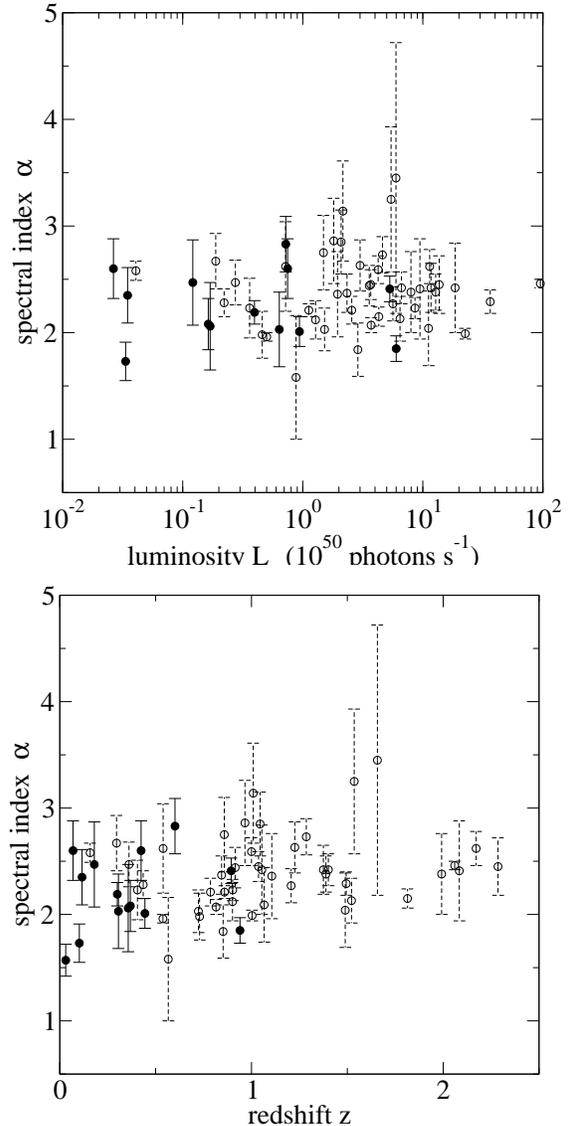

\resizebox{2.9in}{!}
{
\includegraphics{flum.eps}}
\resizebox{2.8in}{!}
{\includegraphics{fz.eps}
}
\vspace{0.1in}
  \caption{\label{fig3}
Spectral index versus luminosity (upper panel) and versus redshift 
(lower panel) for BL Lacs (filled circles/solid
lines) and FSRQs (open circles/dashed lines).}
\end{figure}

If blazars evolve spectrally with redshift, or if their spectral
properties depend on luminosity, then 
the SID calculated from resolved, low-$z$ and/or high-luminosity blazars 
is not representative of the spectral properties of unresolved, 
high-$z$ and/or low-luminosity blazars, and should not be used to
calculate the spectral shape of their collective emission. 
For this reason, we 
test for correlations between 
blazar spectral index and redshift or luminosity. 
Figure \ref{fig3} shows plots of the spectral index versus 
photon luminosity (upper panel) and redshift (lower panel) 
for the 66 3EG confident blazars. 
The (isotropic) photon luminosity $L_p$ (shown in Fig. \ref{fig3} in
units of $10^{50}$ photons/sec) was calculated from the P1234
photon flux $F_{P1234}$ using 
\begin{equation}
L_p = 4\pi F_{P1234} d_L^2 / (1+z)\,.
\end{equation}
Here $d_L$ is the luminosity distance, calculated from $z$ assuming 
a concordance $\Omega_{\rm m}=0.3$, $\Omega_{\Lambda}=0.7$, $H_{0}=70
\mbox{km} \mbox{ s}^{-1} \mbox{Mpc}^{-1}$ cosmology.

Though there is no obvious trend in either panel of
Fig. \ref{fig3}, we apply a formal
non-parametric Spearman test for correlations (e.g. Wall \& Jenkins
2003).  
Given a sample
of $N$ data pairs of variables, the two variables are ranked such that $(X_{i},Y_{i})$ are
the ranks of the variables for the $i$th pair ($1<X_{i}<N$ and 
$1<Y_{i}<N$).  Then,  the Spearman rank correlation coefficient is computed:
\begin{equation}
r_{s} = 1 - 6\frac{\sum_{i=1}^{N} (X_{i}-Y_{i})^2}{N^3-N},
\end{equation}
with range $0<|r_{s}|<1$; a high value indicates a significant
correlation. 
The coefficient values we find for the spectral index/luminosity pairs are
$r_{s}=0.145$ (BL Lacs) and $r_{s}=-0.109$ (FSRQs) and for the
spectral index/redshift pairs are $r_{s}=0.141$ (BL Lacs) and 
$r_{s}=0.238$ (FSRQs). 
In all cases,  such values of $|r_s|$ or smaller occur by 
chance more than $10\%$ of the time.
Therefore, we find no  evidence for correlations between
spectral index and either redshift or luminosity. Accordingly, we find
no evidence for cosmological evolution in the particle acceleration
and $\gamma-$ray emisison 
mechanisms operating in these AGNs, at least in the redshifts under
consideration.

\section{The effect of measurement errors}\label{efer}

We now turn our attention to the reconstruction of the ISID of blazars
using measurements of the spectral indices for individual members of
the blazar population. 
We assume that the ISID 
(probability that a randomly selected blazar has a true spectral index
between $\alpha$ and $\alpha+d\alpha$) can be adequately described by a
Gaussian, 
\begin{equation}\label{gaussISID}
p_{\rm intrinsic}(\alpha)d\alpha = \frac{1}{\sqrt{2\pi \sigma_0}} \exp 
\left[-\frac{(\alpha-\alpha_0)^2}{2\sigma_0^2}\right] d\alpha\,.
\end{equation}
The spread of the ISID (which is equal to $\sigma_0$, and should
 not be confused with the ``error on the mean,'' the
 uncertainty in our knowledge of $\alpha_0$)
is not necessarily well-approximated by the spread
of the distribution of {\em measured} spectral indices. 
In sampling a distribution with significant
errors in individual measurements, the spread of the resulting measured
distribution may be dominated by the
errors, especially if they are comparable with the width of the ISID. 
In this case, if the measured SID
is used to calculate the spectral shape of the collective emission of
unresolved blazars, sources with spectral
indices away from the mean would be overestimated and ultimately lead to
an overestimate of the curvature of the collective emission spectrum. 
It is therefore necessary to determine the degree to which the
measured SID is error-dominated before we can decide whether it is
representative of the ISID.

\begin{figure}
\resizebox{3.0in}{!}
{\includegraphics{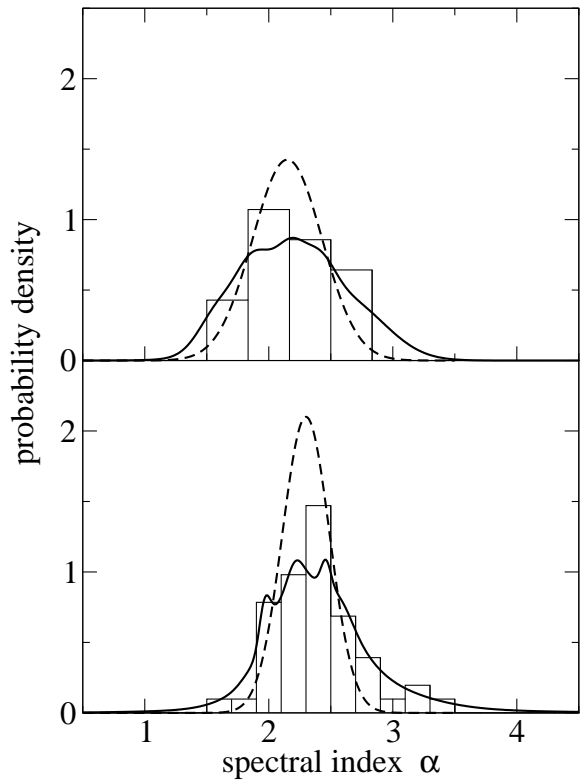}}
  \caption{\label{fig5}
Top:  Histogram of spectral indices of 3EG BL Lacs
  with SS SID (solid) overlayed.  Bottom:  Same for the 3EG FSRQs. 
The dashed lines represent the maximum-likelihood Gaussians
  for the same sets (see \S \ref{LIKEL}).}
\end{figure}

In the formalism implemented by SS96a, the measured SID is reconstructed by
 co-adding all of the probability density functions (assuming Gaussian errors) of individual 
spectral index measurements of blazars in the second EGRET catalog:
\begin{equation}\label{SS}
p_{\rm observed}(\alpha)d\alpha=
\frac{1}{N}\sum_{i=1}^{N}\frac{1}{\sigma_{i}\sqrt{2\pi}}e^{-\frac{(\alpha-\alpha_{i})^2}{2\sigma_{i}^2}}
d\alpha,
\end{equation}
where $\alpha_{i}$ is the spectral index of blazar $i$, $\sigma_{i}$ is the
error in measurement of $\alpha_{i}$, and N is the total number of
blazars.\footnote{SS96a also accounted for the different blazar states
  (flaring/quiescent) by shifting this distribution (toward
  harder/softer indices), but they considered BL Lacs and FSRQs as a
single population.}
Equation (\ref{SS}) is a good representation of the
distribution of {\em measured} spectral indices (see Fig. \ref{fig5}; SS96a
SIDs represented by solid lines) and
has the additional advantage that it does not make
any {\em a priori} assumptions about the shape of the SID, but it could still
be contaminated by measurement errors and as such, might not be a fair
representation of the ISID.  With this concern, we assess the significance of
error in the SID, by performing a simple Monte Carlo test.

We consider separately the samples of
confident  EGRET BL Lacs and FSRQs assuming that the ISID of each subset 
has the form of Eq. (\ref{gaussISID}).  We take the mean of each sample,
$\alpha_{0}$, to be the ``trimmed mean'' (see e.g. Wilcox 1997)
of the sample and the variance $\sigma_0^2$ to be variable. For
 later comparison,
we also calculate, the ``trimmed variance,'' $\sigma_t^2$, of each population. 
We calculate $\alpha_0$ and $\sigma_t^2$
by trimming the top and bottom five percent of the combined populations. 
In this way, we obtain for the BL Lacs: $\alpha_0= 2.20$ and $\sigma_t=0.33$;
and for the FSRQs: $\alpha_0= 2.39$ and $\sigma_t=0.22$. 

\begin{figure}
\resizebox{3.0in}{!}
{\includegraphics{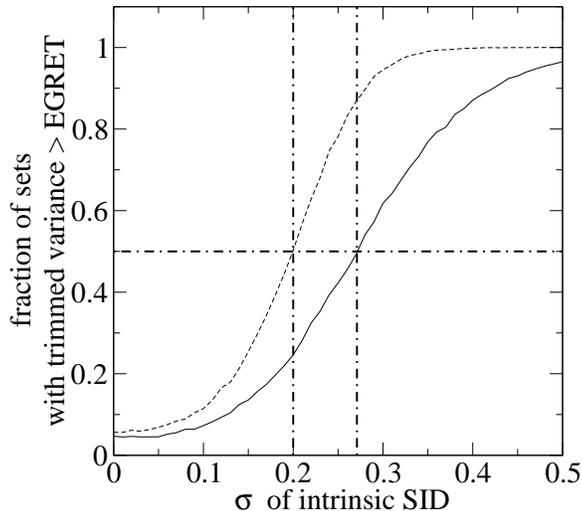}}
  \caption{\label{fig6}
Fraction of sets for which the trimmed variance of the set with
errors is greater than that of the EGRET set without errors for sets with the
    same number of objects as confident BL Lacs (solid line) or FSRQs
    (dotted line). The dot-dashed lines indicate the median ``true
    spread'' of the simulated samples, which is equal to
    $0.27$ for BL Lacs and $0.2$ for FSRQs.}
\end{figure}

We perform a number of ``mock observations''
from each ISID equal to the number of corresponding objects 
EGRET detected, with a randomly chosen
 3EG uncertainty for each object, and calculate the trimmed variances
of each set. 
If some assumed ISID spread, sampled with EGRET uncertainties, more
frequently results in simulated trimmed variances smaller (larger) than that of
the corresponding EGRET dataset, then it is most
likely too small (large) compared with the ISID spread occurring in
nature. The spread of the ISIDs of BL Lacs and FSRQs
occurring in nature should be comparable to the spread for which the
simulated datasets most frequently have trimmed variances
comparable to those of the EGRET datasets. 
Figure \ref{fig6} demonstrates this comparison between the trimmed variances of 
the EGRET set and the simulated sets. The fraction of simulated sets 
with trimmed variance greater than that of the corresponding EGRET set 
is plotted against the spread of the parent ISID.
The median ISID spread is equal to $0.27$ for the BL Lacs
and $0.20$ for FSRQs. If we were to fit a Gaussian to the SS96a 
prescription for the SID, we would obtain a standard
deviation of $0.46$ in the case of BL Lacs and $0.36$ in the case of
FSRQs - almost twice the preferred values of
 the Monte-Carlo analysis. Hence, we
conclude that the SS96a prescription 
does overestimate significantly the spread of the ISID. 

\section{The ISID of EGRET Blazars -
A Likelihood Approach}\label{LIKEL}

Having shown that measurement errors can significantly contaminate the
determination of the ISID,  we now employ a 
likelihood analysis which allows us to explicitly account for
these errors and constrain the parameters of the ISID.
Explicitly accounting for measurement errors is necessary as 
ignoring them may not simply
increase the uncertainty in our estimated parameters but lead to
incorrect parameter inferences (e.g. Loredo 2004). 

Given a set of parameters $x_{i}$, which define a statistical
distribution,
and a dataset $y_{j}$, the probability of $x_{i}$ having certain values
given the data is proportional to the probability of measuring $y_{j}$ given
those values of $x_{i}$ (the {\em likelihood})
times the probability that
$x_{i}$ has those particular values before measurements are taken 
(the {\em prior}):
\begin{equation}
P(x_{i}|y_{j}) \propto P(x_{i}) \times \mathcal{L}(y_{j}|x_{i}).
\end{equation}
If the prior is flat (constant for all values of $x_{i}$),
maximizing the likelihood allows us to determine the most probable values for
the parameters $x_{i}$ (see, e.g., Lee 1989).  

In our analysis, we assume that both the ISID and the errors are
Gaussian. The parameters we wish to calculate are the mean spectral
index, $\alpha_{0}$, and spread, $\sigma_{0}$.  Without error, the likelihood
of measuring a spectral index $\alpha$ given a Gaussian ISID with a mean
of $\alpha_{0}$ and a spread of $\sigma_{0}$ would be
$l = \exp\left[-(\alpha-\alpha_{0})^2/2\sigma_0^2\right]/
\sqrt{2\pi}\sigma_0$. 
To include measurement error, we need to distinguish between the true
spectral index of an object, $\alpha$ (which is unknown
and is therefore a {\em nuisance} parameter over which we need to
marginalize), and its measured spectral
index, $\alpha_j$. Then, the likelihood 
for a single spectral index measurement becomes
\begin{equation}
l_{j} = \int_{-\infty}^{\infty} \!\!d\alpha
\frac{\exp\left[-(\alpha-\alpha_{j})^2/(2\sigma_{j}^2)\right]}
{\sqrt{2\pi}\sigma_{j}}
\frac{ \exp\left[-(\alpha-\alpha_{0})^2/(2\sigma_0^2)\right]}
{\sqrt{2\pi}\sigma_0}
\end{equation}
where the subscript denotes blazar $j$.  If we have $N$ such
independent spectral index measurements, then the overall
likelihood 
becomes
$\mathcal{L} = \prod_{j=1}^{N} l_{j}$ or, 
omitting constant normalization factors, 
\begin{equation}
\mathcal{L} 
= 
\left(
\prod_{j=1}^{N}\frac{1}{\sqrt{\sigma_0^2+\sigma_{j}^2}}
\right)
\exp\left[-\frac{1}{2} \sum_{j=1}^{N}
\frac{(\alpha_{j} - \alpha_{0})^2 }{\sigma_0^2+\sigma_{j}^2}
\right].
\end{equation}
The maximum-likelihood mean and spread of the ISID can be found by maximizing
the likelihood; hence, by simultaneously solving the equations
(for derivation, see Appendix)
\begin{equation}
\alpha_{0} = \left(\sum_{j=1}^{N}
\frac{\alpha_{j}}{\sigma_0^2+
\sigma_{j}^2}\right) 
\left(
\sum_{j=1}^{N}\frac{1}{\sigma_0^2+
\sigma_{j}^2}\right)^{-1},
\end{equation}
and 
\begin{equation}
\sum_{j=1}^{N}\frac{1}{\sigma_0^2+\sigma_{j}^2} 
= \sum_{j=1}^{N}
\frac{(\alpha_{j}-\alpha_{0})^2}{(\sigma_0^2+\sigma_{j}^2)^2}\,.
\end{equation}

\begin{figure}
\resizebox{3.0in}{!}
{\includegraphics{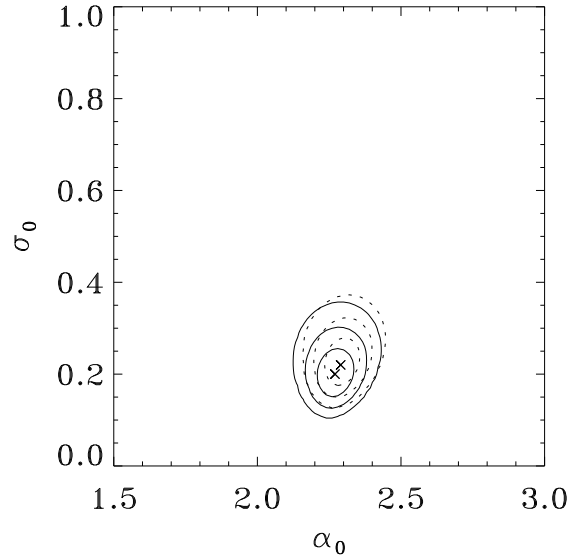}}
  \caption{\label{fig7}
Likelihood function 
contours for the Mattox 2001 (solid lines) and the 3EG
(dashed lines) confident blazar samples. 
The $1\sigma$, $2\sigma$, and $3\sigma$
contours we plot throughout this paper represent
equal-likelihood contours which include $68\%$, $95.5\%$ and $99.7\%$
of the total volume under the likelihood surface.  The x and y axes represent
    the mean ($\alpha_0$) and spread ($\sigma_0$) of the
    ISID respectively. The maximum-likelihood parameters of each ISID
    are denoted by $\times$.}
\end{figure}

In Fig. \ref{fig7}, we plot likelihood contours for the 
confident blazar set of Mattox 2001 (solid lines) and of the 3EG (dashed
lines). 
The maximum-likelihood parameters for the ISID
in the case of the Mattox set are $\alpha_0 = 2.27$ and
$\sigma_0=0.20$, and those for the 3EG set are $\alpha_0=2.29$ 
and $\sigma_0=0.22$. As we can
see in Fig. \ref{fig7}, the likelihood functions of the two sets are 
consistent with each other, which shows that our analysis is
not sensitive to the inclusion or exclusion of a few members.  

\section{BL Lacs vs FSRQS: Do their ISIDs differ?}\label{BvsF}

Gamma-ray--loud BL Lacs and FSRQs have different properties in 
the GeV energy range (e.g. different variability properties, 
Vercellone et al 2004; different mean spectral properties,
Mukherjee et al 1997, Pohl et al 1997; different redshift
distributions and possibly different luminosity functions, 
M\"{u}cke \& Pohl 2000, Dermer 2007). 
It is therefore reasonable to test whether the
maximum-likelihood ISID for the BL Lacs and FSRQs differ
at a statistically significant level.

\begin{figure}
\resizebox{2.9in}{!}
{\includegraphics{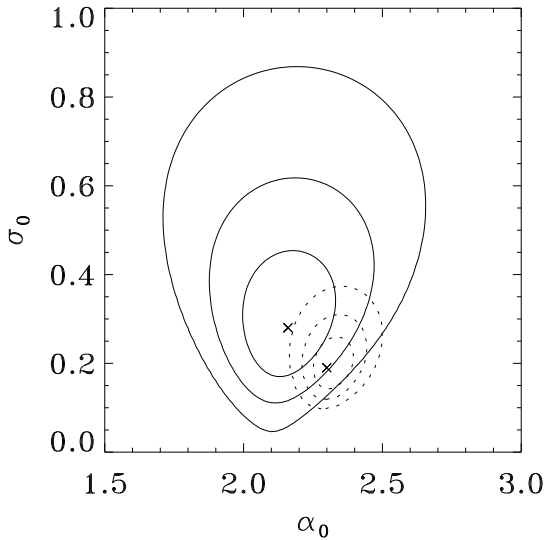}}
  \caption{\label{fig8} 
Likelihood function 
contours (as in Fig. \ref{fig7})
for 3EG BL Lacs (solid lines) and FSRQs (dotted lines).}
\end{figure}

We apply the analysis of \S \ref{LIKEL} to the sets of BL Lacs and
FSRQs of the 3EG catalog. Fig. \ref{fig8} shows the likelihood function
contours for these populations. The ISID parameters are found to be
$\alpha_0=2.15$, $\sigma_0=0.28$ for BL Lacs and $\alpha_0=2.3$,
$\sigma_0=0.19$ 
for FSRQs. The likelihood functions  of the two populations do indicate a
marginal, $1\sigma$ separation. This possible spectral differentiation
between the two populations, if true,  will be confidently confirmed
by GLAST data (see \S \ref{GLAST}). 
The maximum-likelihood $\alpha_0$'s
are close to the trimmed means of each dataset, while the
maximum-likelihood $\sigma_0$'s are almost identical to the median
``true spread'' for each set, as determined by the Monte
Carlo test outlined in \S \ref{efer} indicating the consistency in the results
of the two analyses. 
Our maximum-likelihood Gaussians for BL Lacs and FSRQs 
are overplotted on Fig. \ref{fig5} with the dashed lines.

\section{Blazar Variability and the Measured Spectral Index}\label{VARB}

Though a blazar may spend most of its time in a quiescent state, it can 
undergo periods of flaring during which its flux is significantly
enhanced, up to an order of magnitude  (e.g. McLaughlin
\etal 1996; Mukherjee \etal 1997; Nolan \etal 2003; Vercellone \etal
2004). 
This increase in flux may introduce a detection selection effect for
fainter blazars favoring the flaring state. On the other
hand,  blazars are more likely to be pointed at during
quiescence since they spend more time in the quiescent state.
The spectral index of a blazar is determined using the integrated EGRET maps 
which can include photons from both states, thus 
 complicating the determination of the ISID: it
is unclear whether the spectral index distribution of blazars
determined from EGRET data is representative of blazars in their
flaring or quiescent state. 
This distinction is important because it has been suggested that the
blazar spectrum might change during the flaring state.  If such a 
trend is indeed present and most photons involved in
EGRET blazar detections come from blazars in the flaring state, then 
 the derivation of the SID will be biased towards spectral indices
representative of flaring blazars (as was assumed in SS96a).

Determining whether a blazar is in a flaring or quiescent state during
a certain viewing period is made complicated by low photon statistics.
During the long exposure necessary to detect a blazar, 
the source could be in both
states, and the duty cycle is largely unknown (Vercellone \etal 2004).  
Furthermore, the source 
identification and flux estimation for different viewing periods and
for the cumulative map (from which the spectral index is derived) is done by 
separate statistical processing of each map, 
and as a result, the photon counts
from the individual viewing periods don't add up to those of the cumulative
map. %
With these uncertainties in
mind, we approach the problem using a simple recipe.

In a manner similar to that of Vercellone \etal (2004), we classify each
individual viewing period of a confident AGN (excluding periods for which only
flux upper limits were quoted) as a ``mostly'' flaring period if
the flux of the blazar in the particular viewing period, $F_{\rm VP}$,
is $> \flf \times F_{\rm P1234}$
(where $\flf>1$ is some constant enhancement factor) and if the
error on the measurement of $F_{\rm VP}$ is less than the difference between the
$F_{\rm VP}$  and $F_{\rm P1234}$; otherwise, we classify it as a
``mostly'' quiescent period.  As in Vercellone \etal (2004), 
and in Jorstad et al. (2001), we
consider the conventional case\footnote{Note that Vercellone \etal
  (2004) and 
  Jorstad \etal (2001) compared 
the individual viewing periods with the inverse-uncertainty--weighted mean of
fluxes and upper limits of the individual viewing periods. However,
this weighted flux is more biased towards higher fluxes than the true
long-term average flux of the blazar, as a blazar is preferentially  
detected in a viewing period when it is flaring (Vercellone \etal 2004). Thus,
we systematically identify more ``flaring'' periods than Vercellone \etal (2004) and Jorstad \etal (2001).} of
$\flf=1.5$. We also consider
the case of $\flf=3.75$, chosen as the enhancement factor for which 
$\sim 75\%$ of the
  objects have a flaring fraction below $0.2$.

Once we have classified all of the viewing periods during which a
blazar was detected,
we add all of the photon counts from the flaring periods to get the
number of flaring photons, $N_{f}$, and likewise for the quiescent periods to
get the number of quiescent photons, $N_{q}$.  We then calculate the ``flaring
photon fraction,'' $f=N_{f}/(N_{f}+N_{q})$, for every confident blazar.

\begin{figure}
\resizebox{2.9in}{!}
{\includegraphics{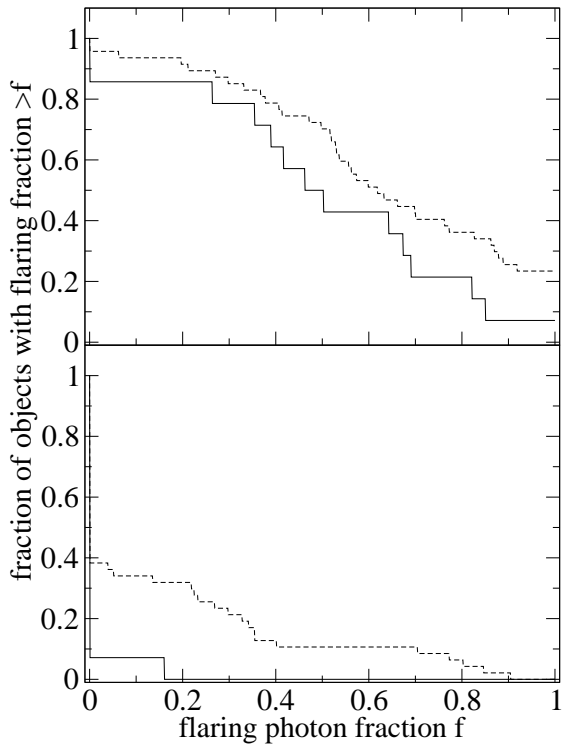}}
\caption{\label{fig2}
Fraction of objects with fraction of flaring photons greater than
$f$ as a function of $f$ for $\flf = 1.5$ (upper panel) and $\flf = 3.75$
(lower panel) for BL Lacs (solid lines) and FSRQs (dotted lines).}
\end{figure}

Figure \ref{fig2} shows the fraction of blazars with a flaring 
fraction greater than $f$ as a function of $f$, for both enhancement factors.
As expected, the flaring photon fraction depends
on our definition of a ``flaring state:''  the fraction of
objects with flaring photon fraction $>f$, is systematically lower for all
values of $f$ for the 
higher value of $\flf$. However, even for $\flf=1.5$,
the median flaring photon fraction is not very high (equal to $0.5$ for BL
Lacs, and $0.6$ for FSRQs, while only $20\%$ of BL Lacs and $40\%$ of FSRQs
have a flaring photon fraction higher than $0.7$, at which point photons predominantly come from a flaring state). 
Therefore, the photons that are used to derive individual blazar spectra
represent a balanced mix of photons originating in flaring and quiescent
states, if not photons that come primarily from the quiescent state (with the
quantitative details of this statement depending on the value of the
enhancement factor). This analysis suggests that even if spectral
indices 
of blazars
do harden considerably during the flaring states for most objects, there is no
indication that the photon budget is overly biased towards one variability
state. Hence, if a SID is determined from
EGRET data, there is no evidence that it would be predominantly representative
of the flaring-state or quiescent-state SID.

Qualitatively, this result also has implications for the blazar duty cycle.
If the time spent flaring  is comparable to the time
spent in quiescence, we would expect the number of flaring photons to be
significantly greater than the number of quiescent photons, yielding a flaring
fraction close to one. This however is not the case even for a
lower value of $\flf$. Therefore, these blazars must have been viewed 
more often in their quiescent state.

\subsection{Is there statistical evidence for spectral index hardening
during flaring?}\label{hardening}

As a first indication for spectral index shift with flaring, 
we can test for a spectral index/flaring photon fraction correlation. 
No obvious trend exists but, as in \S \ref{EVOL}, we also perform
a non-parametric Spearman
rank coefficient test, treating the BL Lacs
and FSRQs separately, and for $\flf=1.5$ and $\flf=3.75$.

With the exception of BL Lacs in the $\flf=3.75$
case, the data are
consistent with uncorrelated variables at the $20\%$ level. 
In the case of $\flf=3.75$, all BL
Lacs but one have $f=0$, which gives a false positive result in the
correlation test. However, when we treat the BL Lacs and the FSRQs as a single
population, the results are again consistent with
uncorrelated variables. A recent systematic reanalysis of EGRET data
for all blazars bright enough for time-resolved spectroscopy by
Nandikotkur et al. (2007) has also yielded similar results despite 
using a radically different approach, as no systematic trend of the spectral
index with changing flux was found.

\begin{figure}
\resizebox{3.0in}{!}
{\includegraphics{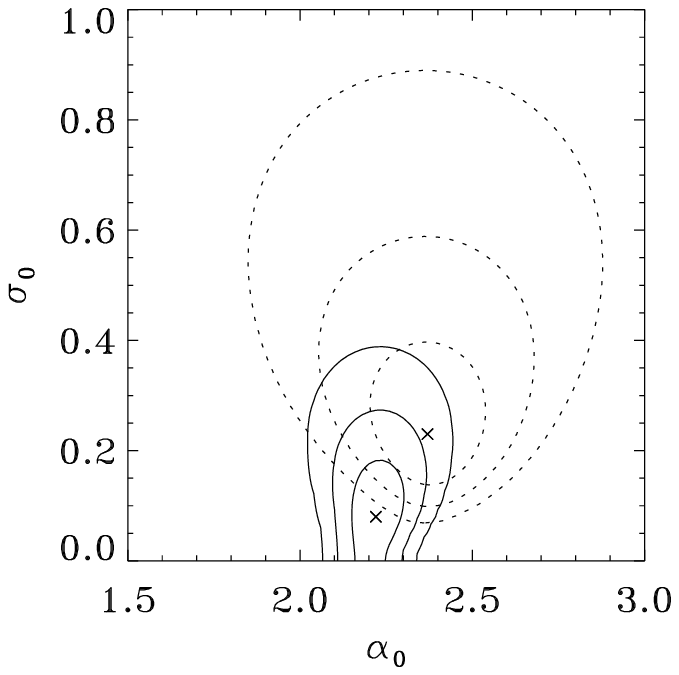}}
\resizebox{3.0in}{!}
{\includegraphics{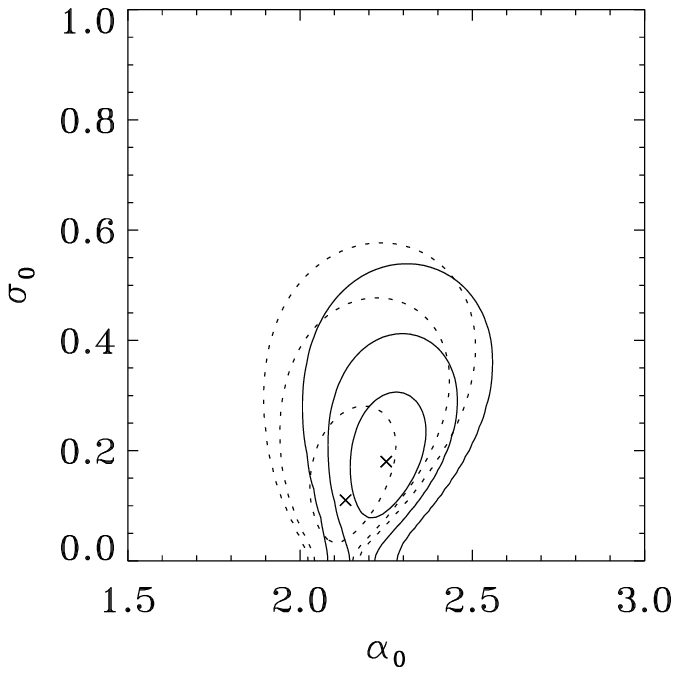}}
  \caption{\label{fig9}
Likelihood function contours (as in Fig. \ref{fig7})  for mostly flaring
    (solid lines) and mostly quiescent (dashed lines) blazars. Upper
    panel: 3EG blazars; lower panel: simulated EGRET dataset.}
\end{figure}

A second method to look for a systematic spectral index shift with
changing flux is to perform the maximum-likelihood analysis of \S
\ref{LIKEL} separately for the populations of ``mostly flaring'' and
``mostly quiescent'' blazars.  The set of ``mostly flaring''
blazars are the 22 EGRET blazars that have a flaring photon
fraction $f>0.7$, for $\flf=1.5$. The set of ``mostly quiescent''
blazars are the 10 EGRET blazars that have flaring photon fractions
$f<0.3$. Likelihood function contours 
for the ISID parameters of these two sets are shown in the upper panel
of Fig. \ref{fig9}. 
The ISIDs of the two sets are fully consistent with each
other, with maximum-likelihood parameters for the ``mostly flaring''
blazars $\alpha_0=2.25$ and $\sigma_0 =0.18$,  and for the ``mostly
quiescent'' blazars $\alpha_0 =2.13$ and $\sigma_0 = 0.11$. If anything, the
maximum-likelihood ISID of mostly flaring objects peaks at slightly {\em
  softer} indices than that of the mostly quiescent sources (a trend that
could be theoretically accomodated if, for example, softer EC emission
overtakes an otherwise harder, SSC dominated, spectrum during flaring; see e.g. B\"ottcher 1999). 

In order to determine whether the EGRET data would be sufficient to reveal
spectral index hardening with flaring if such a trend does exist, we
perform the following test. We assume
that blazar spectral indices do indeed harden with flaring, and that their
ISIDs differ as suggested by SS96a: they are identical Gaussians with
$\sigma_0=0.20$ and means that are determined by shifting the mean of the
combined population, $\alpha_0=2.27$ 
(mean and spread obtained from the maximum-likelihood analysis of the Mattox et al. 2001 set), by $-0.05$
for the flaring blazars and by $+0.2$ for quiescent blazars. 
We then sample this dataset with EGRET uncertainties (as in \S \ref{efer}) and 
create a mock EGRET dataset. 
We apply the same likelihood analysis to these simulated
datasets. Likelihood function contours are plotted in
the lower panel of Fig. \ref{fig9}. 
Although the maximum-likelihood analysis correctly identifies the
flaring population ISID as being harder than the quiescent population
ISID, the separation of the maxima of the likelihood functions has a
marginal $1\sigma$ significance. Additionally, in the EGRET dataset
the measurements of the spectral index do not come from pure
``flaring'' or ``quiescent'' states, and even the definition of a ``flaring''
or ``quiescent'' state is dependent on the selected value of the enhancement
factor $\flf$. Thus, we conclude that although EGRET data do
not seem to indicate a hardening of the spectral index with flaring,
they are not sufficient to exclude such a possibility
either. However, as we will see in the next section, GLAST data will 
settle the issue. 

\section{Prospects for GLAST}\label{GLAST}

With the launch of GLAST later this year, the number statistics of
resolved blazars are expected to improve dramatically. Depending on models
used to determine blazar numbers at lower fluxes, GLAST will detect
between $1,000-10,000$ blazars with flux $>2\times 10^{-9} {\rm \,
  photons \, s^{-1}}$. However, most of these
blazars will have fluxes close to the lower end of GLAST
sensitivity with individual photon statistics and spectral index
uncertainties comparable to those of the EGRET dataset. Thus, the
extent to which the GLAST dataset will improve our understanding of
the issues discussed in this paper is not immediately obvious. 

To assess this question, we construct  
simulated datasets of spectral indices and spectral index uncertainties
for GLAST-detectable blazars. Based on these datasets, we
predict by how much GLAST observations will improve the determination of
statistical properties of the spectral indices of gamma-ray--loud blazars. 
For our study, we use the Dermer (2007) 
gamma-ray luminosity functions for BL Lacs and
FSRQs.  These luminosity functions are more conservative than those of,
e.g., Narumoto \& Totani (2006), Chiang \&
Mukherjee (2001) and SS96a as they predict a {\em smaller} number of
detectable blazars; additionally, 
they treat BL Lacs and FSRQs separately, enabling us to
distinguish between the two populations in our analysis.  If GLAST detects
more blazars than predicted in Dermer (2007) and assumed here, the statistics
will only improve, strengthening the significance of our results. 

In assigning measurement uncertainties to the spectral index of 
each blazar in our simulated datasets, we make use of the strong
anticorrelation of 
spectral index uncertainties in the EGRET dataset 
with the total number of detected P1234 photons.
A single power-law can be fitted to the 3EG data, 
 \begin{equation}\label{Uncert}
\sigma_\alpha = 7.0 \times N_{\rm photon}^{-0.7}\,.
\end{equation} 
We expect that, despite any differences between the {\it Large Area
  Telescope} (LAT) and EGRET instruments, this empirical relation can also
  give a rough approximation of the spectral index uncertainty for blazars
  detected by GLAST.  

\subsection{Spectral separation between BL Lacs and FSRQs}\label{GLASTBvsF}

Using simulated GLAST datasets, we examine 
the possibility that, if BL Lacs and FSRQs are indeed spectrally
distinct populations,  GLAST observations will
be sufficient to separate them with a
significance greater than $3\sigma$.
We assume that the
ISIDs for BL Lacs and FSRQs are the maximum-likelihood Gaussians determined
from EGRET data ($\alpha_0=2.15$ and $\sigma_0=0.28$ for BL Lacs and
$\alpha_0=2.3$ and $\sigma_0=0.19$ for FSRQs).  
We generate the simulated BL Lac and FSRQ datasets by appropriately
sampling these ISIDs with GLAST uncertainties. 
The number of objects in each set and their fluxes 
are derived from Fig. 2 of Dermer (2007). From the flux of each
object, we derive an approximate number of detectable photons,
$N_{{\rm photon},i} = F_i\Delta t A$, 
where $F_i$ is the flux of the particular blazar, $\Delta t$ is the estimated
exposure time (assuming that any given objects will be in the GLAST field of
view for $\sim 20\%$ of a 2 year campaign), and $A$ is the effective
collecting area of the detector ($8,000 {\rm cm^2}$ for GLAST). 
The spectral index uncertainty for each object is  then
 derived from Eq. \ref{Uncert}. 

\begin{figure}
\resizebox{3.0in}{!}
{\plotone{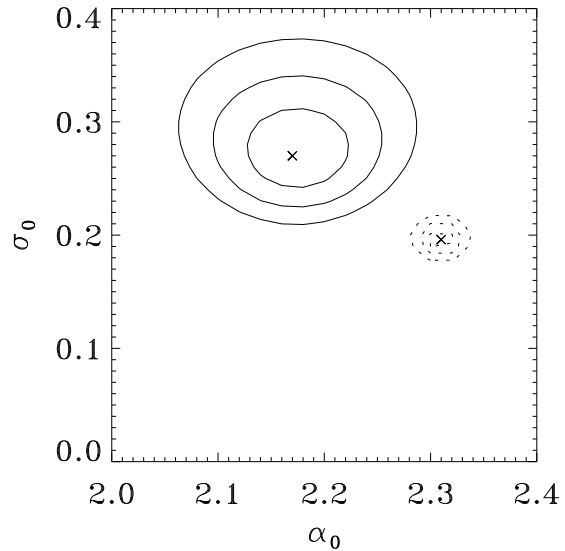}}
  \caption{\label{fig12} Likelihood function 
contours (as in Fig. \ref{fig7}) for simulated GLAST datasets of BL
Lacs (solid lines) and FSRQs (dashed lines)}
\end{figure}

Performing the analysis outlined in \S \ref{LIKEL} on these datasets, we get
the likelihood function contours shown in Fig. \ref{fig12}. The
maximum-likelihood parameters of the ISIDs are $\alpha_0=2.17$ and 
$\sigma_0=0.27$
for the BL Lacs, and $\alpha_0=2.31$ and $\sigma_0=0.2$ for the FSRQs, excellent
approximations of the assumed 
underlying ISID parameters. The separation
between the two populations is clearly established with significance greater
than $3\sigma$.

\subsection{Spectral separation between flaring and quiescent blazars}

Finally, we investigate the possibility 
that GLAST observations will
be sufficient to determine whether flaring and quiescent blazars are indeed
spectrally distinct. We construct simulated GLAST datasets to test two
scenarios: that the ISIDs of flaring and quiescent blazars have the
same shape, but flaring blazars are shifted towards harder indices by 0.25
(as in \S \ref{hardening}); and that flaring and quiescent
blazars have ISIDs with parameters equal to the maximum-likelihood
parameters yielded by the analysis of the true EGRET dataset (note
that in this case the flaring blazars are somewhat {\em softer} than
quiescent blazars). 

As there are many more blazars in this
analysis, we can account for the disparities between the flaring fractions for
bright and faint blazars by calculating (from EGRET data) the fractions of
bright and faint objects that are mostly flaring ($f>0.7$) or mostly quiescent
($f<0.3$); thereby, more carefully determining the numbers of mostly flaring
and mostly quiescent blazars. We assign {\em true} spectral indices as in \S
\ref{hardening} and spectral index uncertainties based on the flux as in \S
\ref{GLASTBvsF}, treating BL Lacs and FSRQs as a single population.

\begin{figure}
\resizebox{3.0in}{!}
{\includegraphics{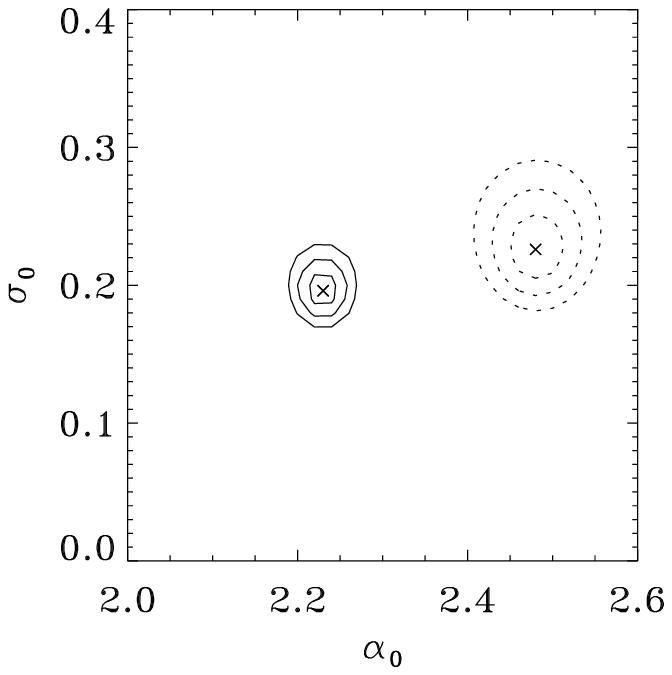}}
\resizebox{3.0in}{!}
{\includegraphics{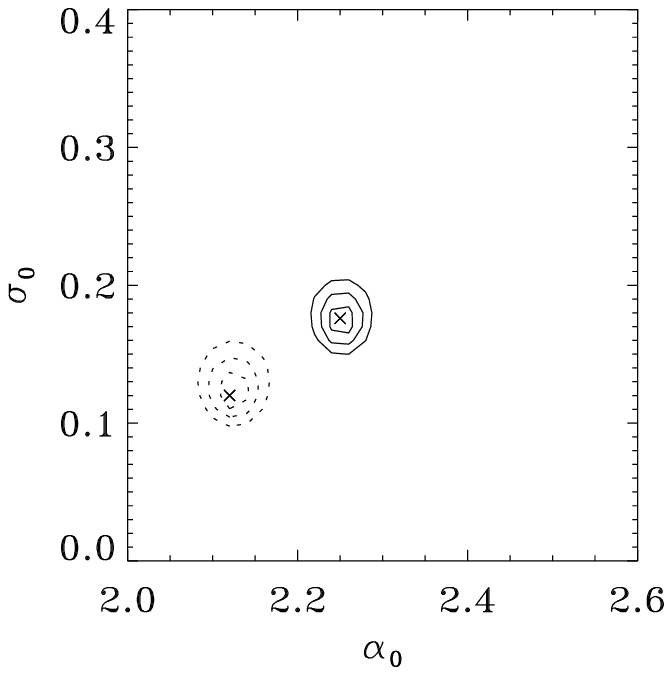}}
  \caption{\label{fig13} Likelihood function contours as in 
Fig. \ref{fig9} for simulated GLAST
    data.  Upper panel: likelihood contours assuming flaring blazars are
  harder; lower panel: likelihood contours assuming flaring and quiescent
  blazars ISID with parameters determined in the analysis of the EGRET
  simulated variability dataset. }
\end{figure}

We then apply the same likelihood analysis to these simulated
datasets. Their likelihood function contours are plotted in
Fig. \ref{fig13}. 
Although the expected partial contamination of any single
state with both flaring and quiescent photons may weaken this result,
likelihood functions for the flaring and the quiescent blazars are
well separated with a significance greater than $3\sigma$ in both scenarios, 
strongly suggesting that the GLAST
data will reveal any systematic change of spectral index with
flaring, if  present. 

\section{Discussion and Summary}\label{Disc}

In this work, we have extensively studied possible trends in the
spectral behavior of EGRET blazars, and have attempted to reconstruct
the blazar ISID from EGRET observations, explicitly accounting for the large
measurement errors arising from low-photon statistics. Additionally, we have
identified still open questions on which progress is expected to be made once
GLAST observations become available, and we have estimated the extent of the
expected improvement. Our conclusions can be
summarized in the following statements:

(a) We have investigated the possibility that source evolution could interfere
with the application of our ISIDs to unresolved blazars. We have 
found no evidence for correlations between spectral index and
redshift or between spectral index and luminosity in EGRET data.  Therefore,
we have concluded that there are no indications for spectral index evolution
in EGRET data.  

(b) We have used a Monte Carlo test to study whether measurement
errors in individual spectral indices severely contaminate the
observed SID. We found that measurement errors can indeed 
have a profound effect and need to be accounted for properly. This will also be
the case for GLAST data, as most of the newly detected blazars will suffer
from similar low-photon statistics as with the EGRET blazars. 
We have performed a
likelihood analysis which explicitly accounts for significant and unequal
errors of measurement, and 
we found a most likely Gaussian ISID for the population
of EGRET blazars with a mean of $2.27$ and a spread of $0.20$,
significantly narrower than the observed SID.  

(c)  We have derived separate ISIDs for BL Lacs and FSRQs finding that FSRQs
are well represented by a Gaussian with mean of $2.3$ and spread of $0.19$,
while the BL Lacs are better represented by a Gaussian with mean of $2.15$ and
spread of $0.28$.  Based on EGRET data, the significance of the spectral
separation of the two populations is marginal.
We determined whether each spectral index 
of the EGRET confident
blazars was more representative of the flaring or quiescent state by
calculating the {\em flaring photon fraction} for each object.  
We performed statistical tests for spectral index hardening and found no
evidence for this effect. We calculated separate
maximum-likelihood ISIDs for mostly flaring and mostly quiescent objects
finding no evidence for a systematic separation of the two populations.
However, we also showed that such a separation can neither be excluded
based on EGRET data. 
This result is consistent with the recent
findings of Nandikotkur et al. (2007) who studied spectral index shifting
during flaring on an object-to-object basis and also found no systematic
trend. Ultimately, the question of hardening will be addressed by GLAST
through more confident measurements of spectral indices and an increased
ability for time-resolved spectroscopy (Dermer \& Dingus 2004) allowing
confident measurements of spectral indices for separate variability 
states of blazars. 

(d) We have made explicit estimates about the improvement of our
statistical analysis once GLAST data become available finding
that, with the detection of many more blazars, GLAST will be able to
separate with significance $>3\sigma$ the ISIDs of BL Lacs and FSRQs, if these
populations are indeed spectrally distinct.
Similarly, we have examined two different scenarios for a possible
systematic spectral shift during flaring finding that in
both cases, GLAST would be able to distinguish between the two ISIDs
at $>3\sigma$. 

The maximum-likelihood analysis presented in this work is therefore a
useful tool in maximizing the information that can be obtained from
the high-uncertainty measurements of spectral indices in
the GeV band, where photon statistics for most objects are very low. 
Additionally, it can provide estimates for {\em both} the
mean and the spread of the underlying, intrinsic distribution(s) of
spectral indices of $\gamma$-ray emitters. Both of these aspects of the
ISID(s) are not only important tests for $\gamma$-ray emission models for
blazars, but also essential inputs in determining the spectral shape
of the diffuse emission from unresolved blazars (Pavlidou \& Venters 2007).
 
\acknowledgements{We gratefully acknowledge enlightening discussions
with Markus B{\"o}ttcher, Chuck Dermer, Brian Fields, Demos Kazanas, Zhaoming Ma, Reshmi
Mukherjee, Angela Olinto, Martin Pohl,  Jennifer Siegal-Gaskins,  
and insightful comments
by Kostas Tassis that improved this paper. T.M.V. would 
also like to thank the members of the
Laboratoire d'AstroParticule et Cosmologie of the Universit{\'e} de Paris 7 for
their hospitality and camaraderie during the final stages of this work.  
This work was supported by the 
Kavli Institute for Cosmological Physics through the grant NSF PHY-0114422. 
T.M.V. was supported by an NSF Graduate Research Fellowship.

\appendix

\section{Derivation and Maximization of Likelihood Function for the
ISID parameters}

As discussed in \S\ref{efer}, the likelihood of observing $N$ spectral
indices
$a_j \, (j=1,...,N)$ with individual measurement uncertainties
$\sigma_j \, (j=1,..,N)$ (where the $\sigma_j$ are assumed Gaussian)
if the ISID is Gaussian with mean $\alpha_0$ and variance $\sigma_0$ is
$\mathcal{L} = \prod_{j=1}^{N} l_{j}$, with

\begin{equation}\label{lj}
l_{j} =
\int_{a=-\infty}^{\infty}\frac{1}{\sigma_{j}\sqrt{2\pi}}
\exp\left[-\frac{(\alpha-\alpha_{j})^{2}}{2\sigma_{j}^{2}}\right]
\frac{1}{\sigma_0\sqrt{2\pi}}\exp\left[-\frac{(\alpha-\alpha_{0})^{2}}
{2\sigma_0^{2}}\right]\,d\alpha\,.
\end{equation}
Defining 
\begin{equation}
A_{j} = \sigma_0^2 + \sigma_{j}^2\,,\,\,\,\,\,
B_{j} = \frac{\sigma_0^2\alpha_{j}+\sigma_{j}^2\alpha_{0}}
{\sigma_0^2+\sigma_{j}^2}\,,\,\,\,\,\,
C_{j} =\frac{\sigma_{j}^2\sigma_0^2(\alpha_{j}-\alpha_{0})^2}
{\sigma_0^2+\sigma_{j}^2}\,,
\end{equation}
we can rewrite the exponentials in Eq. (\ref{lj}) by completing the
square as 
\begin{equation}
\exp\left[-
\frac{(\alpha-\alpha_{j})^{2}}{2\sigma_{j}^{2}} -
 \frac{(\alpha-\alpha_{0})^2}{2\sigma_0^2} \right]
=\exp\left[- 
\frac{A_{j}(\alpha - B_{j})^2 +
C_{j}}{2\sigma_0^2\sigma_{j}^2}
\right].
\end{equation}
Substituting this in Eq. (\ref{lj}) and performing the integration we obtain 
\begin{equation}
l_{j} = 
\exp\left[-\frac{C_{j}}{2\sigma_0^2\sigma_{j}^2}\right]
\frac{1}{\sqrt{2\pi
A_{j}}}\,.
\end{equation}
The likelihood function then becomes
\begin{equation}\label{theL}
\mathcal{L}(\alpha_0,\sigma_0)  =  
\prod_{j=1}^{N}l_{j} =
\left(\prod_{j=1}^{N}\frac{1}{\sqrt{2\pi (\sigma_0^2+\sigma_j^2)}}\right)
\exp\left[
-\frac{1}{2}\sum_{j=1}^N\frac{(\alpha_{j}-\alpha_{0})^2}{\sigma_0^2+\sigma_j^2}
\right]\,.
\end{equation}
Local maxima of this function will satisfy
\begin{equation}\label{partials}
\frac{\partial \mathcal{L}}{\partial \alpha_{0}} = 0, \,\,\,\,
\frac{\partial \mathcal{L}}{\partial \sigma_{0}} = 0.
\end{equation}
Equation (\ref{theL}) implies
\begin{equation}
\frac{\partial \mathcal{L}}{\partial \alpha_{0}} = 
\left(\prod_{j=1}^{N}\frac{1}{\sqrt{2\pi (\sigma_0^2+\sigma_j^2)}}\right)
\exp\left[-\frac{1}{2}\sum_{j=1}^N
\frac{(\alpha_{j}-\alpha_{0})^2}{\sigma_0^2+\sigma_j^2}\right]
\sum_{j=1}^N\frac{\alpha_{j}-\alpha_{0}}{\sigma_0^2+\sigma_j^2}\,,
\end{equation}
therefore the first of Eqs. (\ref{partials}) becomes
\begin{equation}
\alpha_{0} = 
\left(\sum_{j=1}^N
\frac{\alpha_{j}}{\sigma_0^2+\sigma_j^2}\right)
\left(\sum_{j=1}^N\frac{1}{\sigma_0^2+\sigma_j^2}\right)^{-1}\,.
\end{equation}
Similarly, 
\begin{eqnarray}
\frac{\partial \mathcal{L}}{\partial \sigma_0} 
&=&
\exp\left[
-\frac{1}{2}\sum_{j=1}^N\frac{(\alpha_{j}-\alpha_{0})^2}
{\sigma_0^2+\sigma_j^2}
\right] 
\left(
\prod_{j=1}^{N}\sqrt{\frac{1}{2\pi (\sigma_0^2+\sigma_j^2) }}
\right)
\sigma_0
\left[
-\sum_{j=1}^N\frac{1}{\sigma_0^2+\sigma_j^2} + 
\sum_{j=1}^N\frac{(\alpha_{j} - \alpha_{0})^2}{(\sigma_0^2+\sigma_j^2)^2}
\right]\,,
\end{eqnarray}
where we have used the identity 
\begin{equation}
\frac{\partial}{\partial
\sigma_0}\prod_{j=1}^{N}\frac{1}{\sqrt{\sigma_0^2+\sigma_j^2}}=
-\sigma_0\left(\prod_{j=1}^{N}\frac{1}{\sqrt{\sigma_0^2+\sigma_j^2}}\right)
\left(\sum_{j=1}^N\frac{1}{\sigma_0^2+\sigma_j^2}\right) 
\end{equation}
which is straight-forward to prove by induction. Therefore, 
the second of Eqs. (\ref{partials}) becomes 
\begin{equation}
\sum_{j=1}^N\frac{1}{\sigma_0^2+\sigma_j^2} =
\sum_{j=1}^N\frac{(\alpha_{j} - \alpha_{0})^2}{(\sigma_0^2+\sigma_j^2)^2}\,.
\end{equation}

\end{document}